# Effect of magnetic field on transports of charged particles in the weakly ionized plasma with power-law $q$-distributions in nonextensive statistics

Yue Wang and Jiulin Du

*Department of Physics, School of Science, Tianjin University, Tianjin 300072, China*

**Abstract** By using the generalized Boltzmann equation of transport in nonextensive statistics, we study transport properties of the diffusion flux and the heat flux of charged particles in the weakly ionized plasma with the power-law $q$-distributions under the magnetic field. We derive the tensor expressions of diffusion coefficient, thermal diffusion coefficient, mobility and thermal conductivity of electrons and ions in the $q$-distributed plasma under magnetic field. We show that the tensors of the diffusion coefficient, the thermal diffusion coefficient and the thermal conductivity are strongly depend on the $q$-parameters in nonextensive statistics, and so they are generally not the same as those in the magnetized plasma with a Maxwell distribution.

## 1. Introduction

The density and velocity distributions in nonequilibrium complex systems have been observed in astronomy and astrophysics, space plasmas, complex fluids, molecular dynamics, nuclear and atomic physics, condensed matter physics, chemical reaction systems, and biological systems. It has been found that the velocity distributions of the charged particles in the systems with long-range correlations are quite far from the Maxwellian distribution, with long energetic tails [1-2]. Recently, nonextensive statistics proposed by Tsallis has attracted great attention of scientific researchers, a new statistics which can study the complex systems beyond thermodynamic equilibrium [3-4]. Nonextensive statistics is based on two basic postulates [3-5]. The first postulate is the nonextensive entropy $S_q$ of the system, given [3] by

$$S_q = \frac{k_B}{q-1}\left(1 - \sum_{i=1}^{W} p_i^q\right), \quad q \in R, \tag{1}$$

where $k_B$ is Boltzmann constant, $q$ is the nonextensive parameter, and $p_i$ is the probability distribution function at the $i$th state. The second is the experimental measurement of an observable $A$ is defined by the $q$-expectation value,

$$\langle A \rangle_q = \sum_{i=1}^{W} p_i^q a_i, \tag{2}$$

where $a_i$ is the value of $A$ in the $i$th state. Nonextensive statistics was applied to some astrophysical problems [6-10]. The power-law $q$-distribution is a typical distribution under the frame of nonextensive statistics. In astrophysical and space plasmas, the famous $\kappa$-distribution with a non-Maxwellian high-energy tail was first measured by the OGO1 and OGO3 satellites and was applied to explain the data of the low-energy electrons in the magnetosphere [11]. The $q$-distribution is proved to be a natural consequence of the $H$-theorem in nonextensive kinetics, [4]. And many interesting studies on the complex plasmas with power-law distributions in the astrophysics and plasma physics are found by using nonextensive statistics, for instance, plasma oscillations [12], ion acoustic waves and Landau damping [13], instability [14], non-equilibrium plasma with long-range Coulomb interactions [15], and recently transport properties in complex plasmas [16,17].

The $q$-distribution in nonextensive statistics can be used to study the plasma transport processes when the plasma systems are in a stable and non-equilibrium state. In nonextensive statistics, the $q$-parameter for the electromagnetic interactions in the $q$-distributed nonequilibrium plasma is given by the equation [18],

$$k_B \nabla T = e (1-q)(\nabla \varphi - c^{-1} \mathbf{u} \times \mathbf{B}), \tag{3}$$

where $e$ is electron charge, $\varphi$ is the Coulomb potential, $\mathbf{B}$ is the magnetic induction and $c$ is the light speed. The power-law velocity $q$-distribution function for the $\alpha$th plasma component can be expressed as [14,16-17]

$$f_{q,\alpha}^{(0)}(\mathbf{v}) = n_\alpha B_{q,\alpha} \left( \frac{m_\alpha}{2\pi k_B T_\alpha} \right)^{\frac{3}{2}} \left[ 1 - (1-q_\alpha) \frac{m_\alpha \mathbf{v}^2}{2 k_B T_\alpha} \right]^{\frac{1}{1-q_\alpha}}, \tag{4}$$

where $T_\alpha$ is temperature, $n_\alpha$ is particle number density, $m_\alpha$ is particle mass, and $B_{q,\alpha}$ is the $q$-dependent normalized constant given by

$$B_{q,\alpha} = \begin{cases} (1-q_\alpha)^{\frac{1}{2}} (3-q_\alpha)(5-3q_\alpha) \dfrac{\Gamma\left(\dfrac{1}{2} + \dfrac{1}{1-q_\alpha}\right)}{4\Gamma\left(\dfrac{1}{1-q_\alpha}\right)}, & \text{for } q_\alpha \leq 1. \\[2ex] (q_\alpha - 1)^{\frac{3}{2}} \dfrac{\Gamma\left(\dfrac{1}{q_\alpha - 1}\right)}{\Gamma\left(\dfrac{1}{q_\alpha - 1} - \dfrac{3}{2}\right)}, & \text{for } q_\alpha \geq 1. \end{cases} \tag{5}$$

In nonequilibrium complex plasma, the temperature $T$, the concentration $n$ should be considered to be space inhomogeneous, i.e., $T=T(r)$ and $n=n(r)$.

In non-equilibrium plasma, because there are temperature gradient, concentration gradient and potential gradient etc., the transport phenomenon always exist such as the diffusion and the heat conductivity etc. The inter-particle interactions in plasma are bringing about by the collisions between the particles, such as electron-electron collision, electron-ion collision, ion-ion collision, electron-neutral-particle collision and ion-neutral-particle collision. For the weakly ionized plasma, the charged particles are much less than the neutral particles, so there are only electron-neutral -particle collision and ion-neutral-particle collision. Recently, in the weakly ionized plasma with power-law distributions, the diffusion coefficients, the thermal conductivities and the viscosity coefficients were studied respectively[16-17, 21-22]. However, the effect of magnetic field on the plasma transports has not taken into consideration yet. In fact, as we known, the magnetic field plays an very important role in the plasma transport processes and produces changes in the transport coefficients, such as the controlled thermonuclear fusion [23], material processing [24], space propulsion [25] and neutral beam injection for fusion [26]. It is necessary to study the transport coefficients of the power-law distributed plasma under the effect of magnetic field. The plasma with anisotropy due to the presence of an external magnetic field is called the magnetized plasma, such as the plasma in controlled thermonuclear fusion experimental device and in magnetosphere.

In this work, we will study the effect of magnetic field on the transport coefficients of charged particles in the weakly ionized plasma with the *q*-distributions in nonextensive statistics. By studying the diffusion flux and the heat flux in the plasma, we study the transport coefficient tensors in the weakly ionized, the magnetized and the *q*-distributed plasma.

The paper is organized as follows. In Sec.2, we introduce the generalized Boltzmann equation for the weakly ionized plasma with the power-law *q*-distributions in nonextensive statistics. In Sec.3, we calculate the diffusion flux in the plasma with the *q*-distributions and derive the related coefficients, including the diffusion coefficient, the thermal diffusion coefficient and the mobility. In Sec.4, we calculate the heat flux of the plasma in magnetic field with the *q*-distributions and derive the thermal conductivity. In Sec.5, we give the numerical analyses of the effects of the magnetic field and nonextensivity on the transport coefficients. Finally in Sec.6, we give the conclusion.

## 2. The generalized transport equations

In nonextensive statistics, the generalized Boltzmann equation for multi-body system is as following [16, 21-22, 27],

$$\frac{\partial f_\alpha}{\partial t} + \mathbf{v}\cdot\frac{\partial f_\alpha}{\partial \mathbf{r}} + \frac{Q_\alpha}{m_\alpha}\left[\mathbf{E} + \frac{1}{c}(\mathbf{v}\times\mathbf{B})\right]\cdot\frac{\partial f_\alpha}{\partial \mathbf{v}} = C_q(f_\alpha), \qquad (6)$$

where $f_\alpha \equiv f_\alpha(\mathbf{r}, \mathbf{v}, t)$ is a single-particle velocity distribution function at time *t*, velocity **v**, and position **r**; the subscript *α* denotes the electron and the ion, $\alpha = e, i$, respectively, $Q_\alpha$ is the charge of the *α*th plasma component, **E** is the electric field, **B** is the magnetic induction, and *c* is the light speed. Here we use that $\mathbf{r} \equiv (x, y, z)$. $C_q(f_\alpha)$ on the right-hand side is the nonextensive *q*-collision term, which represents the change of the distribution function due to the collisions. For the weakly ionized plasmas, the collisions between the charged particles and the neutral particles mainly determine the transport properties. So we can use the generalized Krook model for the collision term to study the transport properties [16,22]. And we mainly consider the elastic collisions between the charged particles and neutral particles, without the plasma chemistry.

In the generalized Krook model, the collision term for the weakly ionized plasma can be expressed as the following linear term with respect to the distribution function [16,22]

$$C_q(f_\alpha) = -\frac{1}{\tau_\alpha}\left(f_\alpha - f^{(0)}_{q,\alpha}\right) = -\nu_\alpha\left(f_\alpha - f^{(0)}_{q,\alpha}\right), \qquad (7)$$

where $\tau_\alpha$ is the time of transition from a nonequilibrium state to the *q*-distribution state approximately, the mean time of a collision between charged particles and neutral particles. So the $\nu_\alpha$ is the mean collision frequency, $\nu_\alpha = (\tau_\alpha)^{-1}$.

In the first-order approximation of Chapman-Enskog expansion, we can write the velocity distribution function as the following form [28],

$$f_\alpha = f^{(0)}_{q,\alpha} + f^{(1)}_{q,\alpha}, \qquad (8)$$

where $f^{(1)}_{q,\alpha}$ is a first-order small disturbance about the stationary *q*-distribution $f^{(0)}_{q,\alpha}$ with $f^{(1)}_{q,\alpha} \ll f^{(0)}_{q,\alpha}$ [29]. In the case of smooth flow plasma under magnetic field, according to Eqs.(4)-(5), we substitute Eqs.(7)-(8) into Eq.(4), we obtain

$$\left\{\frac{\partial}{\partial t} + \mathbf{v}\cdot\frac{\partial}{\partial \mathbf{r}} + \frac{Q_\alpha}{m_\alpha}\left[\mathbf{E} + \frac{1}{c}(\mathbf{v}\times\mathbf{B})\right]\cdot\frac{\partial}{\partial \mathbf{v}}\right\}\left(f^{(0)}_{q,\alpha} + f^{(1)}_{q,\alpha}\right) = -\nu_\alpha f^{(1)}_{q,\alpha}. \qquad (9)$$

The transport processes are always studied in a stationary state so that $\partial f_\alpha/\partial t = 0$. On the left hand side of Eq.(9), the term $[Q_\alpha/(m_\alpha c)](\mathbf{v}\times\mathbf{B})\cdot \partial f^{(0)}_{q,\alpha}/\partial \mathbf{v} = 0$, the term $\mathbf{v} \cdot \partial f^{(0)}_{q,\alpha}/\partial \mathbf{r}$, the electric field term $(Q_\alpha/m)\mathbf{E} \cdot \partial f^{(0)}_{q,\alpha}/\partial \mathbf{v}$, the magnetic field term $[Q_\alpha/(m_\alpha c)](\mathbf{v}\times\mathbf{B})\cdot \partial f^{(1)}_{q,\alpha}/\partial \mathbf{v}=0$ and the collision term $-\nu_\alpha f^{(1)}_{q,\alpha}$ are retained in the first-order approximation following the textbook [28-32], so Eq.(9) can be simplified as

$$\mathbf{v}\cdot\frac{\partial f^{(0)}_{q,\alpha}}{\partial \mathbf{r}}+\frac{Q_\alpha}{m_\alpha}\mathbf{E}\cdot\frac{\partial f^{(0)}_{q,\alpha}}{\partial \mathbf{v}}+\frac{Q_\alpha}{m_\alpha}\frac{1}{c}(\mathbf{v}\times\mathbf{B})\cdot\frac{\partial f^{(1)}_{q,\alpha}}{\partial \mathbf{v}}=-\nu_\alpha f^{(1)}_{q,\alpha}. \tag{10}$$

Eq.(10) can be rewritten as following

$$\nu_\alpha f^{(1)}_{q,\alpha}+\frac{Q_\alpha}{m_\alpha}\frac{1}{c}(\mathbf{v}\times\mathbf{B})\cdot\frac{\partial f^{(1)}_{q,\alpha}}{\partial \mathbf{v}}=-\frac{Q_\alpha}{m_\alpha}\mathbf{E}\cdot\frac{\partial f^{(0)}_{q,\alpha}}{\partial \mathbf{v}}-\mathbf{v}\cdot\frac{\partial f^{(0)}_{q,\alpha}}{\partial \mathbf{r}}. \tag{11}$$

For further simplification, we assume that the velocity distribution $f^{(1)}_{q,\alpha}$ is axis symmetric and along the direction of diffusion flux (or heat flux) [29], so $f^{(1)}_{q,\alpha}$ can be written as

$$f^{(1)}_{q,\alpha}=\frac{\mathbf{v}\cdot\mathbf{f}^{(1)}_{q,\alpha}}{v}, \tag{12}$$

where $\mathbf{f}^{(1)}_{q,\alpha}$ is a spherically symmetric vector. And then we obtain [30] that

$$\frac{\partial f^{(1)}_{q,\alpha}}{\partial \mathbf{v}}=\frac{\mathbf{v}\mathbf{v}}{v^2}\cdot\frac{\partial \mathbf{f}^{(1)}_{q,\alpha}}{\partial v}+\left(\frac{1}{v}-\frac{\mathbf{v}\mathbf{v}}{v^3}\right)\cdot\mathbf{f}^{(1)}_{q,\alpha}. \tag{13}$$

Substituting Eqs.(12)-(13) into Eq.(11), it can be written as

$$\nu_\alpha\frac{\mathbf{v}\cdot\mathbf{f}^{(1)}_{q,\alpha}}{v}+\frac{Q_\alpha}{m_\alpha c}(\mathbf{v}\times\mathbf{B})\cdot\left[\frac{\mathbf{v}\mathbf{v}}{v^2}\cdot\frac{\partial \mathbf{f}^{(1)}_{q,\alpha}}{\partial v}+\left(\frac{1}{v}-\frac{\mathbf{v}\mathbf{v}}{v^3}\right)\cdot\mathbf{f}^{(1)}_{q,\alpha}\right]=-\frac{Q_\alpha}{m_\alpha}\mathbf{E}\cdot\frac{\partial f^{(0)}_{q,\alpha}}{\partial \mathbf{v}}-\mathbf{v}\cdot\frac{\partial f^{(0)}_{q,\alpha}}{\partial \mathbf{r}}. \tag{14}$$

Considering $(\mathbf{v}\times\mathbf{B})\cdot\mathbf{v}=0$, $(\mathbf{v}\times\mathbf{B})\cdot\mathbf{I}\cdot\mathbf{f}^{(1)}_{q,\alpha}=\mathbf{v}\cdot(\mathbf{B}\times\mathbf{f}^{(1)}_{q,\alpha})$ and $\frac{\partial f^{(1)}_{q,\alpha}}{\partial \mathbf{v}}=\frac{\mathbf{v}}{v}\frac{\partial f^{(1)}_{q,\alpha}}{\partial v}$, then Eq.(14) becomes

$$\nu_\alpha\frac{\mathbf{v}\cdot\mathbf{f}^{(1)}_{q,\alpha}}{v}+\frac{Q_\alpha}{m_\alpha c}\frac{\mathbf{v}}{v}\cdot(\mathbf{B}\times\mathbf{f}^{(1)}_{q,\alpha})=-\frac{Q_\alpha}{m_\alpha}\frac{\mathbf{v}}{v}\cdot\mathbf{E}\frac{\partial f^{(0)}_{q,\alpha}}{\partial v}-\frac{\mathbf{v}}{v}\cdot v\frac{\partial f^{(0)}_{q,\alpha}}{\partial \mathbf{r}}. \tag{15}$$

So we have that

$$\nu_\alpha\mathbf{f}^{(1)}_{q,\alpha}+\frac{Q_\alpha}{m_\alpha c}(\mathbf{B}\times\mathbf{f}^{(1)}_{q,\alpha})=-\frac{Q_\alpha}{m_\alpha}\mathbf{E}\frac{\partial f^{(0)}_{q,\alpha}}{\partial v}-v\frac{\partial f^{(0)}_{q,\alpha}}{\partial \mathbf{r}}. \tag{16}$$

If the direction of the magnetic field is taken along **z** axis, Eq.(16) can be written by the following matrix

$$\begin{pmatrix}\nu_\alpha & -\omega_{Be,\alpha} & 0 \\ \omega_{Be,\alpha} & \nu_\alpha & 0 \\ 0 & 0 & \nu_\alpha\end{pmatrix}\begin{pmatrix}f^{(1)}_{q,\alpha,x} \\ f^{(1)}_{q,\alpha,y} \\ f^{(1)}_{q,\alpha,z}\end{pmatrix}=-\frac{Q_\alpha}{m_\alpha}\begin{pmatrix}E_x \\ E_y \\ E_z\end{pmatrix}\frac{\partial f^{(0)}_{q,\alpha}}{\partial v}-v\begin{pmatrix}\frac{\partial f^{(0)}_{q,\alpha}}{\partial x} \\ \frac{\partial f^{(0)}_{q,\alpha}}{\partial y} \\ \frac{\partial f^{(0)}_{q,\alpha}}{\partial z}\end{pmatrix}, \tag{17}$$

with $\omega_{Be,\alpha}=Q_\alpha B/m_\alpha c$, where $\omega_{Be,\alpha}$ is the Larmor frequency of $\alpha$th plasma component. From Eq.(17), $f^{(1)}_{q,\alpha}(\mathbf{r},\mathbf{v})$ can be solved out as

$$\mathbf{f}^{(1)}_{q,\alpha}=-\mathbf{R}^\alpha\cdot\left(\frac{Q_\alpha}{m_\alpha}\mathbf{E}\frac{\partial f^{(0)}_{q,\alpha}}{\partial v}+v\frac{\partial f^{(0)}_{q,\alpha}}{\partial \mathbf{r}}\right), \tag{18}$$

where $\mathbf{R}^\alpha$ is the inverse matrix of the matrix on the left-hand side of Eq.(17), which can be written [28-30] as

$$\mathbf{R}^\alpha = \begin{pmatrix} \dfrac{v_\alpha}{v_\alpha^2 + \omega_{Be,\alpha}^2} & \dfrac{\omega_{Be,\alpha}}{v_\alpha^2 + \omega_{Be,\alpha}^2} & 0 \\ -\dfrac{\omega_{Be,\alpha}}{v_\alpha^2 + \omega_{Be,\alpha}^2} & \dfrac{v_\alpha}{v_\alpha^2 + \omega_{Be,\alpha}^2} & 0 \\ 0 & 0 & \dfrac{1}{v_\alpha} \end{pmatrix}. \quad (19)$$

Therefore, from Eqs.(8) (12) and (18) the distribution function can be derived as

$$f_\alpha = f_{q,\alpha}^{(0)} - \frac{\mathbf{v}}{v} \cdot \mathbf{R}^\alpha \cdot \left( \frac{Q_\alpha}{m_\alpha} \mathbf{E} \frac{\partial f_{q,\alpha}^{(0)}}{\partial \mathbf{v}} + v \frac{\partial f_{q,\alpha}^{(0)}}{\partial \mathbf{r}} \right). \quad (20)$$

## 3. The diffusion flux and the transport coefficients

If the temperature, density and potential of the plasma are space inhomogeneous, the charged particles will naturally move from the regions of high temperature, high density and high potential to the regions of low temperature, low density and low potential energy. Because the magnetic field makes the plasma anisotropic, the diffusion flux for $\alpha$th component is written [30, 33-34] as

$$\mathbf{J}_{m,\alpha} = -\boldsymbol{\alpha} \cdot \nabla \mu_\alpha - \boldsymbol{\beta} \cdot \nabla T_\alpha + m_\alpha n_\alpha \mathbf{b} \cdot \mathbf{f}, \quad (21)$$

where $\mu_\alpha$ is the chemical potential of $\alpha$th component, $\boldsymbol{\alpha}$, $\boldsymbol{\beta}$ and $\mathbf{b}$ are the related coefficients in the tensor form of the diffusion flux to the thermodynamic forces, respectively: $\boldsymbol{\alpha}$ is proportional to the diffusion coefficient, $\boldsymbol{\beta}$ is proportional to the thermal diffusion coefficient describing the Soret effect, and $\mathbf{b}$ is the mobility due to the external force $\mathbf{f}$. The chemical potential is a function of the pressure $p_\alpha$, the density $n_\alpha$ and the temperature $T_\alpha$, Further, Eq.(21) can be written as

$$\mathbf{J}_{m,\alpha} = m_\alpha n_\alpha Q_\alpha \mathbf{b}_{q,\alpha} \cdot \mathbf{E} - m_\alpha \mathbf{D}_{q,\alpha} \cdot \nabla n_\alpha - \frac{m_\alpha}{T_\alpha} \left( k_T \mathbf{D}_{q,\alpha} \right) \cdot \nabla T_\alpha - \frac{m_\alpha}{p_\alpha} \left( k_p \mathbf{D}_{q,\alpha} \right) \cdot \nabla p_\alpha, \quad (22)$$

if we denote that $\mathbf{f} = \mathbf{E} Q_\alpha$,

$$\mathbf{D}_{q,\alpha} = \frac{\boldsymbol{\alpha}}{m_\alpha} \left( \frac{\partial \mu_\alpha}{\partial n_\alpha} \right)_{T_\alpha, P_\alpha}, \quad (23)$$

$$k_T \mathbf{D}_{q,\alpha} = \frac{T_\alpha}{m_\alpha} \left[ \boldsymbol{\alpha} \left( \frac{\partial \mu_\alpha}{\partial T_\alpha} \right)_{n_\alpha, P_\alpha} + \boldsymbol{\beta} \right], \quad (24)$$

and

$$k_p \mathbf{D}_{q,\alpha} = \frac{p_\alpha \boldsymbol{\alpha}}{m_\alpha} \left( \frac{\partial \mu_\alpha}{\partial p_\alpha} \right)_{n_\alpha, T_\alpha}. \quad (25)$$

where $\mathbf{D}_{q,\alpha}$ is the diffusion coefficient tensor, $k_T \mathbf{D}_{q,\alpha}$ is the thermal diffusion coefficient tensor, and $k_p \mathbf{D}_{q,\alpha}$ is the barodiffusion coefficient tensor [34]. The barodiffusion coefficient is taken into consideration only when there is a considerable pressure gradient in the fluid. Plasma barodiffusion in inertial-confinement-fusion implosions was applied to observed yield anomalies in thermonuclear fuel mixtures [35], barodiffusion of oxygen, sulphur and silicon caused the stratified layer at the core–mantle boundary in the Earth's outer core [36], and the barodiffusion constant existed in the case of a mixture with a small relative difference in molecular weights of the components [37].

The microscopic representation of the diffusion flux with the distribution function $f_\alpha(\mathbf{r}, \mathbf{v}, t)$ of the $\alpha$th component is

$$\mathbf{J}_{m,\alpha} = m_\alpha \iiint \mathbf{v} f_\alpha(\mathbf{r}, \mathbf{v}, t) d\mathbf{v}. \tag{26}$$

Substituting Eq.(20) into Eq.(26), we have that

$$\mathbf{J}_{m,\alpha} = m_\alpha \iiint \mathbf{v} \left[ f_{q,\alpha}^{(0)} - \mathbf{v} \cdot \mathbf{R}^\alpha \cdot \frac{\partial f_{q,\alpha}^{(0)}}{\partial \mathbf{r}} - \frac{Q_\alpha}{m_\alpha} \frac{\mathbf{v}}{v} \cdot \mathbf{R}^\alpha \cdot \mathbf{E} \frac{\partial f_{q,\alpha}^{(0)}}{\partial v} \right] d\mathbf{v}. \tag{27}$$

On the right-hand side of Eq.(27), the first term of the integral is given by

$$m_\alpha \iiint \mathbf{v} f_{q,\alpha}^{(0)} d\mathbf{v} = m_\alpha n_\alpha B_{q,\alpha} \left( \frac{m_\alpha}{2\pi k_B T_\alpha} \right)^{\frac{3}{2}} \iiint \mathbf{v} \left[ 1 - (1-q_\alpha) \frac{m_\alpha \mathbf{v}^2}{2 k_B T_\alpha} \right]^{\frac{1}{1-q_\alpha}} d\mathbf{v}. \tag{28}$$

The integral is zero because the integrand is an odd function about $\mathbf{v}$. The second term of the integrand in Eq.(27) is

$$-m_\alpha \iiint \mathbf{v}\mathbf{v} \cdot \mathbf{R}^\alpha \cdot \frac{\partial f_{q,\alpha}^{(0)}}{\partial \mathbf{r}} d\mathbf{v}$$

$$= -m_\alpha \iiint \mathbf{v}\mathbf{v} \cdot \mathbf{R}^\alpha \cdot \left\{ \frac{1}{n_\alpha} \frac{\partial n_\alpha}{\partial \mathbf{r}} + \frac{m_\alpha \mathbf{v}^2}{2 k_B T_\alpha^2} \left[ 1 - (1-q_\alpha) \frac{m_\alpha \mathbf{v}^2}{2 k_B T_\alpha} \right]^{-1} \frac{\partial T_\alpha}{\partial \mathbf{r}} - \frac{3}{2} \frac{1}{T_\alpha} \frac{\partial T_\alpha}{\partial \mathbf{r}} \right\} f_{q,\alpha}^{(0)} d\mathbf{v}. \tag{29}$$

On the right-hand side of Eq.(29), the first term of the integral is given (see Appendix) by

$$-\frac{m_\alpha}{n_\alpha} \iiint \mathbf{v}\mathbf{v} \cdot \mathbf{R}^\alpha \cdot \frac{\partial n_\alpha}{\partial \mathbf{r}} f_{q,\alpha}^{(0)} d\mathbf{v} = -\frac{2 k_B T_\alpha}{(7-5q_\alpha)} \mathbf{R}^\alpha \cdot \nabla n_\alpha, \quad 0 < q_\alpha < \frac{7}{5}. \tag{30}$$

Comparing Eq.(30) with Eq.(22), we find the diffusion coefficient tensor of charged particles in the weakly ionized and magnetized plasma with the $q$-distributions,

$$\mathbf{D}_{q,\alpha} = \frac{2 k_B T_\alpha \mathbf{R}^\alpha}{m_\alpha (7-5q_\alpha)}$$

$$= \frac{2 k_B T_\alpha}{m_\alpha (7-5q_\alpha)} \begin{pmatrix} \frac{v_\alpha}{v_\alpha^2 + \omega_{Be,\alpha}^2} & \frac{\omega_{Be,\alpha}}{v_\alpha^2 + \omega_{Be,\alpha}^2} & 0 \\ -\frac{\omega_{Be,\alpha}}{v_\alpha^2 + \omega_{Be,\alpha}^2} & \frac{v_\alpha}{v_\alpha^2 + \omega_{Be,\alpha}^2} & 0 \\ 0 & 0 & \frac{1}{v_\alpha} \end{pmatrix}, \quad 0 < q_\alpha < \frac{7}{5}. \tag{31}$$

On the right-hand side of Eq.(29), the second and the third terms of the integrand are given by

$$-m_\alpha \iiint \mathbf{v}\mathbf{v} \cdot \left\{ \frac{m_\alpha \mathbf{v}^2}{2 k_B T_\alpha^2} \left[ 1 - (1-q_\alpha) \frac{m_\alpha \mathbf{v}^2}{2 k_B T_\alpha} \right]^{-1} \mathbf{R}^\alpha \cdot \frac{\partial T_\alpha}{\partial \mathbf{r}} - \frac{3}{2} \frac{1}{T_\alpha} \mathbf{R}^\alpha \cdot \frac{\partial T_\alpha}{\partial \mathbf{r}} \right\} f_{q,\alpha}^{(0)} d\mathbf{v}$$

$$= -\frac{2 n_\alpha k_B}{(7-5q_\alpha)} \mathbf{R}^\alpha \cdot \nabla T_\alpha, \quad 0 < q_\alpha < \frac{7}{5}. \tag{32}$$

Comparing Eq.(32) with Eq.(22), we obtain the thermal diffusion coefficient tensor of charged particles in the plasma with the $q$-distributions,

$$k_T \mathbf{D}_{q,\alpha} = \frac{2n_\alpha k_B T_\alpha \mathbf{R}^\alpha}{(7-5q_\alpha)m_\alpha}$$

$$= \frac{2n_\alpha k_B T_\alpha}{(7-5q_\alpha)m_\alpha} \begin{pmatrix} \frac{\nu_\alpha}{\nu_\alpha^2+\omega_{Be,\alpha}^2} & \frac{\omega_{Be,\alpha}}{\nu_\alpha^2+\omega_{Be,\alpha}^2} & 0 \\ -\frac{\omega_{Be,\alpha}}{\nu_\alpha^2+\omega_{Be,\alpha}^2} & \frac{\nu_\alpha}{\nu_\alpha^2+\omega_{Be,\alpha}^2} & 0 \\ 0 & 0 & \frac{1}{\nu_\alpha} \end{pmatrix}, \quad 0 < q_\alpha < \frac{7}{5}. \tag{33}$$

The third term of the integral in Eq.(27) is

$$-Q_\alpha \mathbf{R}^\alpha \cdot \mathbf{E} \frac{1}{3} \iiint v \frac{\partial f_{q,\alpha}^{(0)}}{\partial v} d\mathbf{v} = n_\alpha Q_\alpha \mathbf{R}^\alpha \cdot \mathbf{E}, \quad 0 < q_\alpha < \frac{5}{3}. \tag{34}$$

Comparing Eq.(34) with Eq.(22), we find the mobility tensor of charged particles in the weakly ionized and magnetized plasma with the $q$-distributions,

$$\mathbf{b}_{q,\alpha} = \frac{\mathbf{R}^\alpha}{m_\alpha} = \frac{1}{m_\alpha} \begin{pmatrix} \frac{\nu_\alpha}{\nu_\alpha^2+\omega_{Be,\alpha}^2} & \frac{\omega_{Be,\alpha}}{\nu_\alpha^2+\omega_{Be,\alpha}^2} & 0 \\ -\frac{\omega_{Be,\alpha}}{\nu_\alpha^2+\omega_{Be,\alpha}^2} & \frac{\nu_\alpha}{\nu_\alpha^2+\omega_{Be,\alpha}^2} & 0 \\ 0 & 0 & \frac{1}{\nu_\alpha} \end{pmatrix}. \tag{35}$$

According to Eq.(30), (32) and (34) with the consistency of parameter $q_\alpha$, Eq.(27) is finally written as

$$\mathbf{J}_{m,\alpha} = n_\alpha Q_\alpha \mathbf{R}^\alpha \cdot \mathbf{E} - \frac{2k_B T_\alpha}{(7-5q_\alpha)} \mathbf{R}^\alpha \cdot \nabla n_\alpha - \frac{2n_\alpha k_B}{(7-5q_\alpha)} \mathbf{R}^\alpha \cdot \nabla T_\alpha, \quad 0 < q_\alpha < \frac{7}{5}. \tag{36}$$

If we take the limit $\mathbf{B} \to 0$ and $q_\alpha \to 1$, Eq.(36) recovers the diffusion flux in the same plasma without magnetic field and with a Maxwellian distribution [32,34,38],

$$\mathbf{J}_{m,\alpha} = \frac{n_\alpha Q_\alpha}{\nu_\alpha} \mathbf{E} - \frac{k_B T_\alpha}{\nu_\alpha} \nabla n_\alpha - \frac{n_\alpha k_B}{\nu_\alpha} \nabla T_\alpha. \tag{37}$$

**4. The heat flux and the thermal conductivity**

The heat flux describes the energy diffusion of charged particles from a higher temperature, higher concentration and higher potential region to a lower temperature, lower concentration and lower potential region if the temperature, concentration and potential in the plasma are space inhomogeneous. Magnetic field makes the plasma anisotropic, so the generalized Fourier's law is written [30, 33-34] as

$$\mathbf{J}_{q,\alpha} = -T_\alpha \boldsymbol{\beta} \cdot \nabla \mu_\alpha - \boldsymbol{\gamma} \cdot \nabla T_\alpha + \mu_\alpha \mathbf{J}_{m,\alpha} + \mathbf{c} \cdot \mathbf{f}, \tag{38}$$

where $\mathbf{J}_{q,\alpha}$ is the heat flux of $\alpha$th component; $\mathbf{c}$ is the electrothermal effect tensor, also called Benedicks effect, describing a heat flux caused by the electric field [39-40]. replacing $\nabla \mu_\alpha$ by $\mathbf{J}_{m,\alpha}$ and $\nabla T_\alpha$, and then substituting Eq.(21) into Eq.(38), we obtain that

$$\mathbf{J}_{q,\alpha} = \left(\mu_\alpha \mathbf{1} + T_\alpha \boldsymbol{\beta} \cdot \boldsymbol{\alpha}^{-1}\right) \cdot \mathbf{J}_{m,\alpha} - \boldsymbol{\kappa}_{q,\alpha} \cdot \nabla T_\alpha + Q_\alpha \left(\mathbf{c} - m_\alpha n_\alpha T_\alpha \boldsymbol{\beta} \cdot \boldsymbol{\alpha}^{-1} \cdot \mathbf{b}\right) \cdot \mathbf{E}, \tag{39}$$

where

$$\boldsymbol{\kappa}_{q,\alpha} = \left(\boldsymbol{\gamma} - T\boldsymbol{\beta} \cdot \boldsymbol{\alpha}^{-1} \cdot \boldsymbol{\beta}\right) \tag{40}$$

is the thermal conductivity of $\alpha$th component. From Eq.(22) and Eq.(39), it is shown that the density gradient can not only produce diffusion flux, but also produce heat flux, which is called Dufour effect [39].

We now write the microscopic representation of heat flux with the distribution function $f_\alpha(\mathbf{r}, \mathbf{v}, t)$ of $\alpha$th component:

$$\mathbf{J}_{q,\alpha} = \frac{1}{2} m_\alpha \iiint v^2 \mathbf{v} f_\alpha(\mathbf{r}, \mathbf{v}, t) d\mathbf{v}. \tag{41}$$

Substituting Eq.(20) into Eq.(41), we have that

$$\mathbf{J}_{q,\alpha} = \frac{1}{2} m_\alpha \iiint v^2 \mathbf{v} \left[ f_{q,\alpha}^{(0)} - \mathbf{v} \cdot \mathbf{R}^\alpha \cdot \frac{\partial f_{q,\alpha}^{(0)}}{\partial \mathbf{r}} - \frac{Q_\alpha}{m_\alpha} \frac{\mathbf{v}}{v} \cdot \mathbf{R}^\alpha \cdot \mathbf{E} \frac{\partial f_{q,\alpha}^{(0)}}{\partial v} \right] d\mathbf{v}. \tag{42}$$

On the right-hand side of Eq.(42), the first term of the integrand is given by

$$\frac{1}{2} m_\alpha \iiint v^2 \mathbf{v} f_{q,\alpha}^{(0)} d\mathbf{v} = \frac{1}{2} m_\alpha n_\alpha B_{q,\alpha} \left(\frac{m_\alpha}{2\pi k_B T_\alpha}\right)^{\frac{3}{2}} \iiint v^2 \mathbf{v} \left[1 - (1-q_\alpha)\frac{m_\alpha v^2}{2k_B T_\alpha}\right]^{\frac{1}{1-q_\alpha}} d\mathbf{v}. \tag{43}$$

The integrand in Eq.(43) is zero because the integrand is an odd function about $\mathbf{v}$. The second term of the integrand in Eq.(42) is that

$$-\frac{1}{2} m_\alpha \iiint v^2 \mathbf{v} \mathbf{v} \cdot \mathbf{R}^\alpha \cdot \frac{\partial f_{q,\alpha}^{(0)}}{\partial \mathbf{r}} d\mathbf{v}$$

$$= -\frac{m_\alpha}{2} \iiint v^2 \mathbf{v} \mathbf{v} \cdot \mathbf{R}^\alpha \cdot \left\{\frac{1}{n_\alpha} \frac{\partial n_\alpha}{\partial \mathbf{r}} + \frac{m_\alpha v^2}{2k_B T_\alpha^2}\left[1-(1-q_\alpha)\frac{m_\alpha v^2}{2k_B T_\alpha}\right]^{-1} \frac{\partial T_\alpha}{\partial \mathbf{r}} - \frac{3}{2}\frac{1}{T_\alpha}\frac{\partial T_\alpha}{\partial \mathbf{r}}\right\} f_{q,\alpha}^{(0)} d\mathbf{v}. \tag{44}$$

On the right-hand side in Eq.(44), the first term of the integrand in Eq.(44) is

$$-\frac{m_\alpha}{2n_\alpha} \iiint v^2 \mathbf{v} \mathbf{v} \cdot \mathbf{R}^\alpha \cdot \frac{\partial n_\alpha}{\partial \mathbf{r}} f_{q,\alpha}^{(0)} d\mathbf{v} = -\frac{10(k_B T_\alpha)^2}{(7-5q_\alpha)(9-7q_\alpha)m_\alpha} \mathbf{R}^\alpha \cdot \nabla n_\alpha, \quad 0 < q_\alpha < \frac{9}{7}, \tag{45}$$

and the second term of the integrand is

$$-\frac{m_\alpha}{2} \iiint v^2 \mathbf{v} \mathbf{v} \cdot \left\{\frac{m_\alpha v^2}{2k_B T_\alpha^2}\left[1-(1-q_\alpha)\frac{m_\alpha v^2}{2k_B T_\alpha}\right]^{-1} \mathbf{R}^\alpha \cdot \frac{\partial T_\alpha}{\partial \mathbf{r}} - \frac{3}{2}\frac{1}{T_\alpha}\mathbf{R}^\alpha \cdot \frac{\partial T_\alpha}{\partial \mathbf{r}}\right\} f_{q,\alpha}^{(0)} d\mathbf{v}$$

$$= -\frac{20 n_\alpha k_B^2 T_\alpha}{(9-7q_\alpha)(7-5q_\alpha)m_\alpha} \mathbf{R}^\alpha \cdot \nabla T_\alpha, \quad 0 < q_\alpha < \frac{9}{7}. \tag{46}$$

The third term of the integrand in Eq.(42) is

$$-\frac{Q_\alpha}{2} \mathbf{R}^\alpha \cdot \mathbf{E} \frac{1}{3} \iiint v^3 \frac{\partial f_{q,\alpha}^{(0)}}{\partial v} d\mathbf{v} = \frac{5 Q_\alpha n_\alpha k_B T_\alpha}{(7-5q_\alpha)m_\alpha} \mathbf{R}^\alpha \cdot \mathbf{E}, \quad 0 < q_\alpha < \frac{7}{5}. \tag{47}$$

So Eq.(42) is finally calculated as

$$\mathbf{J}_{q,\alpha} = \frac{5Q_\alpha n_\alpha k_B T_\alpha \mathbf{R}^\alpha \cdot \mathbf{E}}{(7-5q_\alpha)m_\alpha} - \frac{10(k_B T_\alpha)^2 \mathbf{R}^\alpha \cdot \nabla n_\alpha}{(7-5q_\alpha)(9-7q_\alpha)m_\alpha} - \frac{20 n_\alpha k_B^2 T_\alpha \mathbf{R}^\alpha \cdot \nabla T_\alpha}{(9-7q_\alpha)(7-5q_\alpha)m_\alpha}, \quad 0 < q_\alpha < \frac{9}{7}. \tag{48}$$

Using Eq.(36), Eq.(48) can be written as

$$\mathbf{J}_{q,\alpha} = \frac{5 k_B T_\alpha \mathbf{J}_{m,\alpha}}{(9-7q_\alpha)m_\alpha} - \frac{10 n_\alpha k_B^2 T_\alpha \mathbf{R}^\alpha \cdot \nabla T_\alpha}{(9-7q_\alpha)(7-5q_\alpha)m_\alpha} + \frac{10(1-q_\alpha)k_B T_\alpha n_\alpha Q_\alpha \mathbf{R}^\alpha \cdot \mathbf{E}}{(7-5q_\alpha)(9-7q_\alpha)m_\alpha}, \quad 0 < q_\alpha < \frac{9}{7}. \tag{49}$$

Comparing Eq.(49) with Eq.(39), we can find the thermal conductivity tensor:

$$\boldsymbol{\kappa}_{q,\alpha} = \frac{10 n_\alpha k_B^2 T_\alpha \mathbf{R}^\alpha}{(9-7q_\alpha)(7-5q_\alpha)m_\alpha}$$

$$= \frac{10 n_\alpha k_B^2 T_\alpha}{(9-7q_\alpha)(7-5q_\alpha)m_\alpha} \begin{pmatrix} \dfrac{v_\alpha}{v_\alpha^2 + \omega_{Be,\alpha}^2} & \dfrac{\omega_{Be,\alpha}}{v_\alpha^2 + \omega_{Be,\alpha}^2} & 0 \\ -\dfrac{\omega_{Be,\alpha}}{v_\alpha^2 + \omega_{Be,\alpha}^2} & \dfrac{v_\alpha}{v_\alpha^2 + \omega_{Be,\alpha}^2} & 0 \\ 0 & 0 & \dfrac{1}{v_\alpha} \end{pmatrix}, \quad 0 < q_\alpha < \frac{9}{7}. \tag{50}$$

It is clear that when we take the limit $\mathbf{B} \to 0$ and $q_\alpha \to 1$, Eq.(50) recovers the expression of the thermal conductivity in the plasma with a Maxwellian distribution and without magnetic field [32, 34, 38],

$$\kappa_{1,\alpha} = \frac{5 n_\alpha k_B^2 T_\alpha}{2 m_\alpha v_\alpha}. \tag{51}$$

Eqs.(31), (33) and (50) show us clearly that the diffusion coefficient tensor, the thermal diffusion coefficient tensor and the thermal conductivity tensor all depend strongly on the $q$-parameter different from unity. Therefore, the transport processes in the plasma with the power-law velocity $q$-distribution should be quite different from those with a Maxwellian distribution. However, Eq.(35) show that the mobility is independent of the $q$-parameter and so it is still the same as the plasma with a Maxwellian distribution.

For the magnetic-field-free plasma, we can take $\omega_{Be,\alpha}=0$ in Eq.(19),

$$\mathbf{R}^\alpha = \begin{pmatrix} v_\alpha^{-1} & 0 & 0 \\ 0 & v_\alpha^{-1} & 0 \\ 0 & 0 & v_\alpha^{-1} \end{pmatrix} \tag{52}$$

so the transport coefficient tensors in Eqs.(31), (33), (35) and (50) all become scalars, namely, there are only the following four coefficients:

$$\mathbf{D}_{q,\alpha} = D_{q,\alpha,\parallel} = \frac{2 k_B T_\alpha}{(7-5q_\alpha)m_\alpha v_\alpha}, \quad 0 < q_\alpha < \frac{7}{5}, \tag{53}$$

$$k_T \mathbf{D}_{q,\alpha} = k_T D_{q,\alpha,\parallel} = \frac{2 n_\alpha k_B T_\alpha}{(7-5q_\alpha)m_\alpha v_\alpha}, \quad 0 < q_\alpha < \frac{7}{5}, \tag{54}$$

$$\mathbf{b}_{q,\alpha} = b_{q,\alpha,\parallel} = \frac{1}{m_\alpha v_\alpha}, \tag{55}$$

and

$$\kappa_{q,\alpha} = \kappa_{q,\alpha,\parallel} = \frac{10 n_\alpha k_B^2 T_\alpha}{(9-7q_\alpha)(7-5q_\alpha) m_\alpha \nu_\alpha}, \quad 0 < q_\alpha < \frac{9}{7}. \tag{56}$$

They are the previous results derived in the plasma without the magnetic field [22].

## 5. Numerical analyses for the transport coefficients

In order to clearly show the effects of magnetic field and nonextensivity on the transport coefficients of charged particles in the weakly ionized and magnetized plasma with power-law velocity $q$-distributions, we make numerical analyses of the diffusion coefficient, the thermal diffusion coefficient and the thermal conductivity, respectively.

We take the diffusion coefficient under magnetic field as an example. From Eq.(22) and Eq.(31), we derived that

$$J_{m,\alpha,x} = -D_{q,\alpha,\perp} \frac{\partial n_\alpha}{\partial x} - D_{q,\alpha,T} \frac{\partial n_\alpha}{\partial y}, \tag{57}$$

$$J_{m,\alpha,y} = -D_{q,\alpha,\perp} \frac{\partial n_\alpha}{\partial y} + D_{q,\alpha,T} \frac{\partial n_\alpha}{\partial x}, \tag{58}$$

and

$$J_{m,\alpha,z} = -D_{q,\alpha,\parallel} \frac{\partial n_\alpha}{\partial z}, \tag{59}$$

where

$$D_{q,\alpha,\perp} = \frac{\nu_\alpha}{\left(\nu_\alpha^2 + \omega_{Be,\alpha}^2\right)} D_{q,\alpha,\parallel}, \quad D_{q,\alpha,T} = \frac{\omega_{Be,\alpha}}{\left(\nu_\alpha^2 + \omega_{Be,\alpha}^2\right)} D_{q,\alpha,\parallel}, \tag{60}$$

$$D_{q,\alpha,\parallel} = \frac{2 k_B T_\alpha}{(7-5q_\alpha)\nu_\alpha}, \quad 0 < q_\alpha < \frac{7}{5}. \tag{61}$$

From Eq.(60)-(61), we find that the longitudinal diffusion coefficient $D_{q,\alpha\parallel}$ (along the direction of magnetic field) is not affected by magnetic field, but the transverse diffusion coefficient $D_{q,\alpha\perp}$ and the Hall diffusion coefficient $D_{q,\alpha,T}$ (perpendicular to the direction of magnetic field) are affected by the influence of magnetic field. If without the magnetic field, we can take the limit $\omega_{Be} \to 0$, then the Hall diffusion coefficient $D_{q,\alpha,T}$ disappears and the transverse diffusion coefficient $D_{q,\alpha,\perp}$ becomes the same as the longitudinal diffusion coefficient $D_{q,\alpha\parallel}$. The same rule can be applied to the thermal diffusion coefficient, the mobility and the thermal conductivity. So we can also obtain, respectively, the longitudinal thermal diffusion coefficient $k_T D_{q,\alpha\parallel}$, the transverse thermal diffusion coefficient $k_T D_{q,\alpha,\perp}$, the Hall thermal diffusion coefficient $k_T D_{q,\alpha,T}$, the longitudinal thermal conductivity $\kappa_{q,\alpha,\parallel}$, the transverse thermal conductivity $\kappa_{q,\alpha,\perp}$ and the Hall thermal conductivity coefficient $\kappa_{q,\alpha,T}$.

In order to show the effects more clearly of the $q$-parameter on these transport coefficients, from (31), (33), and (50) we can write the relations as:

$$\begin{cases} \dfrac{D_{q,\alpha,\perp}}{D_{1,\alpha}} = \dfrac{2}{(7-5q_\alpha)} \left[1 + \left(\dfrac{\omega_{Be,\alpha}^2}{\nu_\alpha^2}\right)\right]^{-1}, \\[2mm] \dfrac{D_{q,\alpha,T}}{D_{1,\alpha}} = \dfrac{2\omega_{Be,\alpha}}{(7-5q_\alpha)\nu_\alpha} \left[1 + \left(\dfrac{\omega_{Be,\alpha}^2}{\nu_\alpha^2}\right)\right]^{-1}, \quad 0 < q_\alpha < \dfrac{7}{5}, \\[2mm] \dfrac{D_{q,\alpha,\parallel}}{D_{1,\alpha}} = \dfrac{2}{(7-5q_\alpha)}, \end{cases} \tag{62}$$

$$\begin{cases} \dfrac{k_T D_{q,\alpha,\perp}}{k_T D_{1,\alpha}} = \dfrac{2}{(7-5q_\alpha)}\left[1+\left(\dfrac{\omega_{Be,\alpha}^2}{v_\alpha^2}\right)\right]^{-1}, \\ \dfrac{k_T D_{q,\alpha,T}}{k_T D_{1,\alpha}} = \dfrac{2\omega_{Be,\alpha}}{(7-5q_\alpha)v_\alpha}\left[1+\left(\dfrac{\omega_{Be,\alpha}^2}{v_\alpha^2}\right)\right]^{-1}, \quad 0<q_\alpha<\dfrac{7}{5}, \\ \dfrac{k_T D_{q,\alpha,\parallel}}{k_T D_{1,\alpha}} = \dfrac{2}{(7-5q_\alpha)}, \end{cases} \quad (63)$$

and

$$\begin{cases} \dfrac{\kappa_{q,\alpha,\perp}}{\kappa_{1,\alpha}} = \dfrac{4}{(9-7q_\alpha)(7-5q_\alpha)}\left[1+\left(\dfrac{\omega_{Be,\alpha}^2}{v_\alpha^2}\right)\right]^{-1}, \\ \dfrac{\kappa_{q,\alpha,T}}{\kappa_{1,\alpha}} = \dfrac{4\omega_{Be,\alpha}}{(9-7q_\alpha)(7-5q_\alpha)v_\alpha}\left[1+\left(\dfrac{\omega_{Be,\alpha}^2}{v_\alpha^2}\right)\right]^{-1}, \quad 0<q_\alpha<\dfrac{9}{7}, \\ \dfrac{\kappa_{q,\alpha,\parallel}}{\kappa_{1,\alpha}} = \dfrac{4}{(9-7q_\alpha)(7-5q_\alpha)}, \end{cases} \quad (64)$$

where denominator on the left-hand side of each equation in Eqs.(62)-(64) is the transport coefficient in the plasma with a Maxwellian distribution and without magnetic field.

On the basis of the above equations Eqs.(62)-(64), the numerical analyses have been made to show the roles of the magnetic field and nonextensive parameters in the transport coefficients: the diffusion coefficient (or the thermal diffusion coefficient) and the thermal conductivity, relative to those in the case of the plasma with a Maxwellian distribution and without magnetic field. In Figs.1-4, respectively, we give the curves of these transport coefficients for three different values of the $q$-parameters, $q_\alpha$=0.2, 1.0, and 1.2, where $D_{q,\alpha,\perp}/D_{1,\alpha}$, $D_{q,\alpha,T}/D_{1,\alpha}$, $\kappa_{q,\alpha,\perp}/\kappa_{1,\alpha}$ and $\kappa_{q,\alpha,T}/\kappa_{1,\alpha}$, are as the ordinate axis and $\omega_{Be}/v_\alpha$ is as the abscissa axis in these figures, respectively.

In Figs.1 and 3, it is shown that the transverse diffusion coefficient (or the transverse thermal diffusion coefficient) and the transverse thermal conductivity all decrease monotonously as the magnetic field increases. But in Figs.2 and 4, it is shown that as the magnetic field increases, the Hall diffusion coefficient (or the Hall thermal diffusion coefficient) and the Hall thermal conductivity all start to increase, reach a maximum and then decrease.

The reason why the $D_{q,\alpha,T}/D_{1,\alpha}$ and $\kappa_{q,\alpha,T}/\kappa_{1,\alpha}$ appear the maximum values in a specific magnitude of the magnetic field can be explained by the fact called the Hall effect: the charged particles deflect and get together under Lorentz force introduced by the magnetic field and thus generate an additional electric field; and then the Lorentz force on the charged particles is gradually offset by the additional electric field force produced, so the diffusion flux and the heat flux which are perpendicular to the direction of the magnetic field and velocity are weakened.

In all the Figs.1-4, we show clearly that the $q$-parameter different from unity plays a significant role in these transport processes, and so all these transport coefficients should be generally different from those in the plasma with a Maxwellian distribution. In the present cases, it is shown that they increase generally as the $q$-parameter increases.

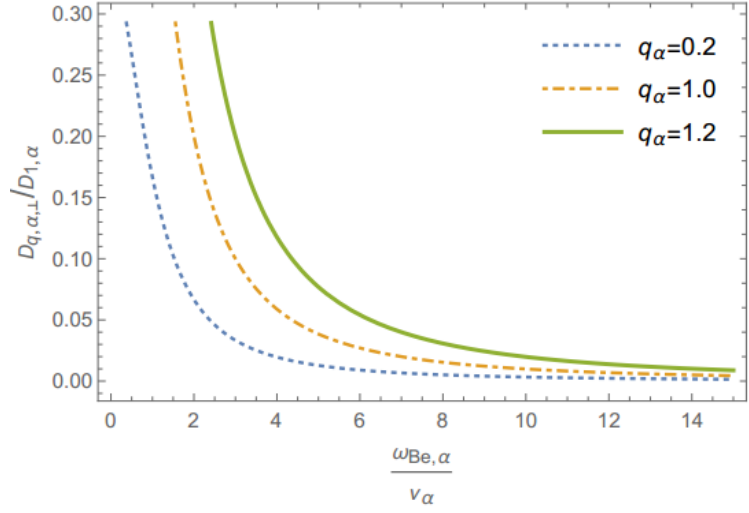

Fig.1 The role of magnetic field in $D_{q,\,\alpha,\,\perp}$

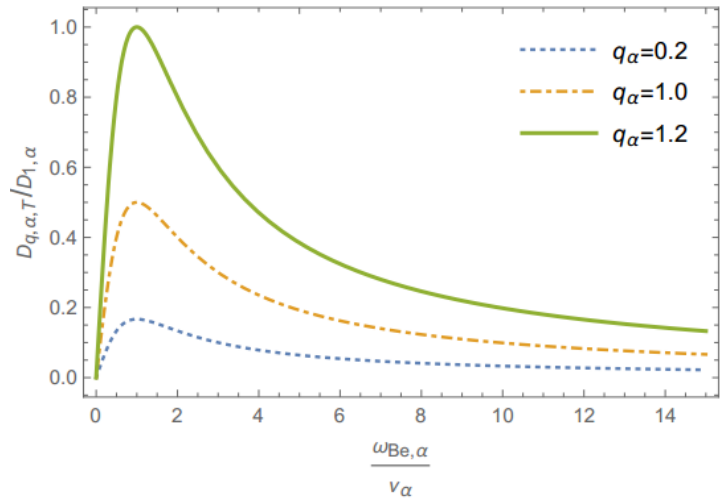

Fig.2 The role of magnetic field in $D_{q,\,\alpha,\,T}$

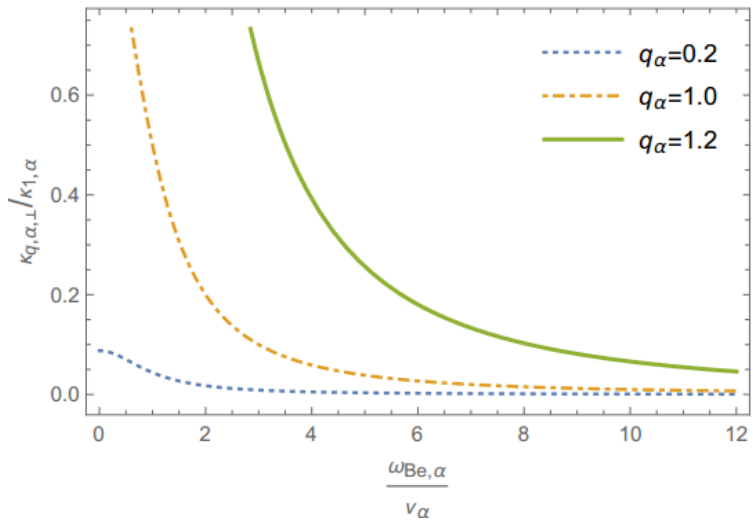

Fig.3 The role of magnetic field in $\kappa_{q,\,\alpha,\,\perp}$

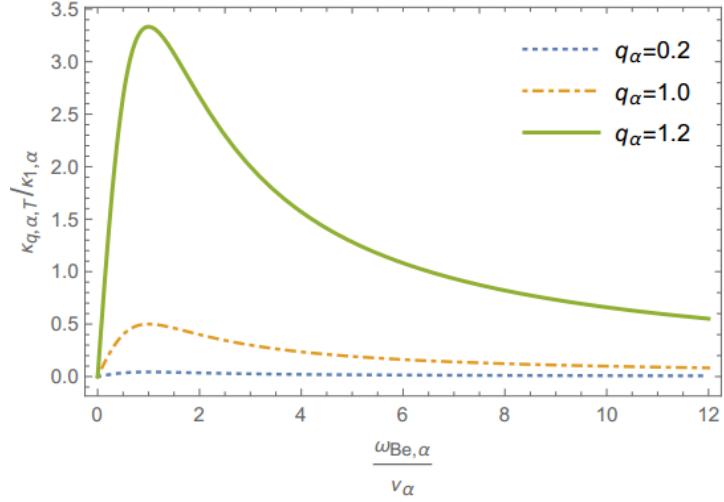

Fig.4 The role of magnetic field in $\kappa_{q,\,\alpha,\,T}$

## 6. Conclusion and discussion

In conclusion, we have studied the effect of the magnetic field on the transport coefficients of charged particles in the weakly ionized plasma with power-law velocity $q$-distributions in nonextensive statistics. From the microscopic expressions of the diffusion flux and the heat flux of the charged particles in the plasma, we derived the expressions of the diffusion coefficient, the thermal diffusion coefficient tensor, the mobility tensor and the thermal conductivity tensor, given by Eq.(31), Eq.(33), Eq.(35) and Eq.(50), respectively. These new transport coefficients not only depend strongly on the magnetic field, but also on the $q$-parameter in the power-law $q$-distributed plasma. We find that that the diffusion coefficient tensor, the thermal diffusion coefficient tensor and the thermal conductivity tensor are all significantly modified by the $q$-parameter different from unity. Therefore, the transport processes in the plasma with the power-law velocity $q$-distribution should be quite different from those with a Maxwellian distribution. However, the mobility is independent of the $q$-distribution and so it is still the same as the plasma with a Maxwellian distribution.

Finally, as we well known, the existence of magnetic field still makes the $q$-distributed plasma anisotropy and so the transport coefficients in the plasma are still tensors.

## Acknowledgement

This work is supported by the National Natural Science Foundation of China under Grant No. 11775156.

## Appendix

Eq.(31) is calculated as follows.

$$-\frac{m_\alpha}{n_\alpha}\iiint \mathbf{vv}\cdot\mathbf{R}^\alpha\cdot\frac{\partial n_\alpha}{\partial \mathbf{r}}f_{q,\alpha}^{(0)}d\mathbf{v} = -\frac{m_\alpha}{3n_\alpha}\mathbf{R}^\alpha\cdot\frac{\partial n_\alpha}{\partial \mathbf{r}}\iiint v^2 f_{q,\alpha}^{(0)}d\mathbf{v}$$

$$= -\frac{4\pi m_\alpha}{3n_\alpha}\mathbf{R}^\alpha\cdot\frac{\partial n_\alpha}{\partial \mathbf{r}}\int_0^{+\infty} v^4 f_{q,\alpha}^{(0)}d\mathbf{v}$$

$$= -\frac{4\pi m_\alpha}{3}\mathbf{R}^\alpha\cdot\frac{\partial n_\alpha}{\partial \mathbf{r}}B_q\left(\frac{m_\alpha}{2\pi k_B T_\alpha}\right)^{\frac{3}{2}}\int_0^{+\infty} v^4\left[1-(1-q_\alpha)\frac{m_\alpha v^2}{2k_B T_\alpha}\right]^{\frac{1}{1-q_\alpha}}d\mathbf{v}. \qquad (A.1)$$

For $q_\alpha > 1$, the integrals in Eq.(A.1) are calculated as

$$\int_0^{+\infty} v^4 \left[1 - (1-q_\alpha)\frac{m_\alpha v^2}{2k_B T_\alpha}\right]^{\frac{1}{1-q_\alpha}} dv$$
$$= 3\sqrt{\frac{\pi}{2}} \left[\frac{m_\alpha (q_\alpha-1)}{k_B T_\alpha}\right]^{-\frac{5}{2}} \frac{\Gamma\left(-\frac{5}{2}+\frac{1}{q_\alpha-1}\right)}{\Gamma\left(\frac{1}{q_\alpha-1}\right)}, \quad 1 < q_\alpha < \frac{7}{5}.$$ 
(A.2)

And therefore we obtain that

$$-\frac{m_\alpha}{n_\alpha}\iiint \mathbf{vv}\cdot\mathbf{R}^a \cdot \frac{\partial n_\alpha}{\partial \mathbf{r}} f_{q,\alpha}^{(0)} d\mathbf{v} = -\frac{2k_B T_\alpha}{(7-5q_\alpha)} \mathbf{R}^a \cdot \nabla n_\alpha, \quad 1 < q_\alpha < \frac{7}{5}. \tag{A.3}$$

For $0 < q_\alpha < 1$, there is a thermal cutoff. The integrals in (A.2) are calculated as

$$\int_0^{\sqrt{\frac{2k_B T_\alpha}{m_\alpha(1-q_\alpha)}}} v^4 \left[1 - (1-q_\alpha)\frac{m_\alpha v^2}{2k_B T_\alpha}\right]^{\frac{1}{1-q_\alpha}} dv$$
$$= 3\sqrt{\frac{\pi}{2}} \left[\frac{m_\alpha (1-q_\alpha)}{k_B T_\alpha}\right]^{-\frac{5}{2}} \frac{\Gamma\left(1+\frac{1}{1-q_\alpha}\right)}{\Gamma\left(\frac{7}{2}+\frac{1}{1-q_\alpha}\right)}, \quad 0 < q_\alpha < 1.$$
(A.4)

And therefore we obtain that

$$-\frac{m_\alpha}{n_\alpha}\iiint \mathbf{vv}\cdot\mathbf{R}^a \cdot \frac{\partial n_\alpha}{\partial \mathbf{r}} f_{q,\alpha}^{(0)} d\mathbf{v} = -\frac{2k_B T_\alpha}{(7-5q_\alpha)} \mathbf{R}^a \cdot \nabla n_\alpha, 0 < q_\alpha < 1. \tag{A.5}$$